\documentclass[12pt,a4paper]{article}
\usepackage{indentfirst}
\usepackage{graphicx}
\usepackage{epsfig}

\begin{document}
\begin{center}
  {\bf  System Size and Centrality Dependence of the Electric Charge}
\end{center}

\vspace{-0.8 cm}

\begin{center}
  {\bf Correlations in A+A and p+p Collisions at the SPS Energies}
\end{center}

\begin{center}
  {P. Christakoglou\footnote{{\bf e - mail : Panos.Christakoglou@cern.ch}}, 
A. Petridis, M. Vassiliou}
\end{center}

\vspace{-1 cm}

\begin{center}
  {University of Athens}
\end{center}

\vspace{-1 cm}
\begin{center}
  {for the NA49 Collaboration}
\end{center}

\textbf{Abstract}

The Balance Function analysis method was developed in order to study the long 
range correlations in pseudo-rapidity of charged particles. The final results on 
p+p, C+C, Si+Si and centrality selected Pb+Pb collisions at $\sqrt{s_{NN}} = 17.2$ 
GeV and the preliminary data at $\sqrt{s_{NN}} = 8.8$ GeV are presented. The 
width of the Balance Function decreases with increasing system size and centrality 
of the collisions. This could suggest a delayed hadronization scenario.

\vspace{0.2 cm}

%%%%%%%%%%%%%%%%%%%%%%%%%%%%%%%%%%%%
\noindent \textbf{1. Introduction}
%%%%%%%%%%%%%%%%%%%%%%%%%%%%%%%%%%%%

Experiments at RHIC, the CERN SPS and the Brookhaven AGS used heavy ion 
collisions as a tool to investigate the possible creation of a deconfined 
phase of matter, the quark-gluon plasma (QGP) \cite{QGP}. This was done by 
studying a variety of characteristics such as particle yields and spectra 
\cite{QM}. Among these characteristics, the study of correlations and 
fluctuations is expected to provide additional information on the reaction 
mechanism of high energy nuclear collisions \cite{Stock99}. 

An important measure of correlations, the Balance Function (BF), was introduced 
by Bass, Danielewicz and Pratt \cite{Pratt}. The first results on the BF were 
obtained for Au+Au collisions by the STAR collaboration at RHIC \cite{STAR}.

In this contribution we concentrate on the study of the pseudo - rapidity 
correlation of charged particles in p+p, C+C, Si+Si and centrality selected 
Pb+Pb collisions at a center-of-mass energy of $\sqrt{s_{NN}} = 17.2$ GeV 
\cite{BF_NA49}. We also present our preliminary results at $\sqrt{s_{NN}} = 8.8$ 
GeV. The data were obtained with the NA49 detector at the CERN SPS \cite{na49_nim}.

The Balance Function (BF) is defined as the difference of the correlation functions 
in pseudo - rapidity $\eta$ of oppositely charged particles and the correlation 
functions of same charged particles, normalized to the total number of particles 
\cite{Pratt}. The BF is given by the following equation:

\vspace{-0.5 cm}

\begin{center}
\begin{equation}
B(\Delta \eta) = \frac{1}{2} \Big[ \frac{N_{+-}(\Delta \eta) -
N_{--}(\Delta \eta)}{N_{-}} + 
\frac{N_{-+}(\Delta \eta) - N_{++}(\Delta \eta)}{N_{+}}  \Big].
\label{BF_DEF2}
\end{equation}
\end{center}

The most interesting property of the BF is its width. Early stage hadronization 
results in a broad BF, while late stage hadronization leads to a narrower 
distribution. The BFs probe the dynamics of charge - anticharge particle pairs 
by quantifying the degree to which the charges are correlated in rapidity space 
\cite{Pratt}. The width of the BF can be characterized by the weighted average 
$\langle \Delta \eta \rangle$ which is calculated by the following formula:

\vspace{-1 cm}

\begin{center}
\begin{equation}
\langle \Delta \eta \rangle = \sum_{i=0}^k{(B_i \cdot \Delta \eta _i)}/\sum_{i=0}^k{B_i},
\label{width}
\end{equation}
\end{center}

\noindent where \emph{i} is the bin number of the BF histogram. 

\vspace{0.2 cm}
%%%%%%%%%%%%%%%%%%%%%%%%%%%%%%%%%%%%%%%%%%%%
\noindent \textbf{2. Data Analysis and Results}
%%%%%%%%%%%%%%%%%%%%%%%%%%%%%%%%%%%%%%%%%%%%

The data sets that have been used in this analysis come from p+p, C+C, Si+Si and 
Pb+Pb collisions at $\sqrt{s_{NN}} = 17.2$ and $\sqrt{s_{NN}} = 8.8$ GeV. For Pb+Pb 
interactions data with both central and minimum bias trigger have been analyzed in 
order to study the centrality dependence of the BF.

For $\sqrt{s_{NN}} = 17.2$ GeV only tracks which passed the NA49 acceptance filter 
\cite{Jacek} and satisfy the following criteria: $0.005 < p_T < 1.5$ GeV/c and 
$ 2.6 < \eta < 5.0$ are used in the analysis. For $\sqrt{s_{NN}} = 8.8$ GeV the 
acceptance filter was not applied and the tracks had to satisfy the criteria: 
$0.005 < p_T < 1.5$ GeV/c and $ 1.6 < \eta < 4.0$.

The mixing procedure, called shuffling, was used to provide an estimate of the maximum 
possible value of the width of the BF while retaining the constraint of charge conservation.

Finally, in order to further investigate the system size and centrality dependence 
of the BF, we generated events coming from p+p, non elastic C+C and Si+Si collisions 
as well as centrality selected  Pb+Pb interactions at $\sqrt{s_{NN}} = 17.2$ GeV using 
the HIJING event generator \cite{Hijing}.

\begin{figure}[ht]
\begin{center}
\epsfig{angle=270,file=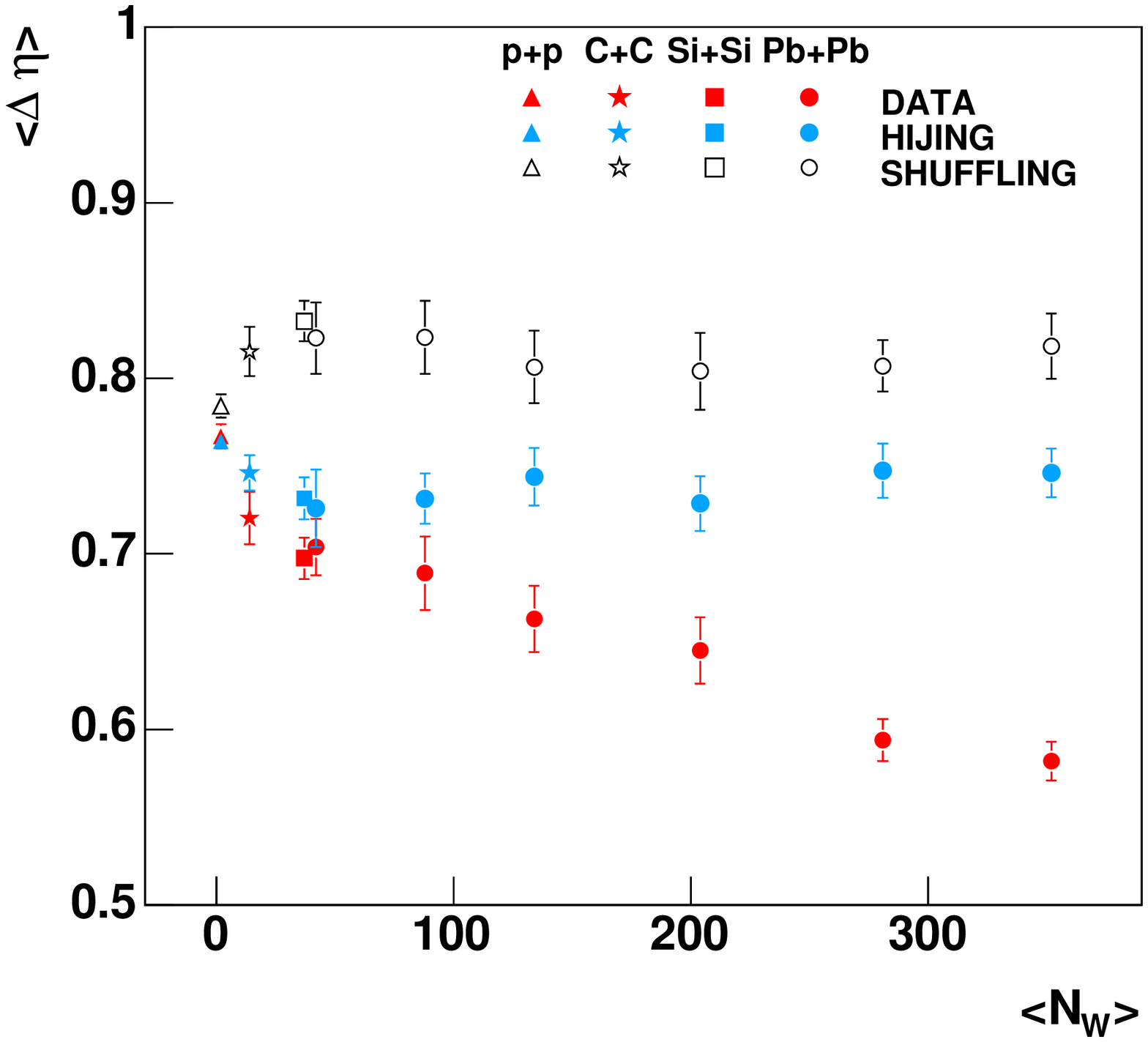,width=6.5cm}
\hspace{0.1 cm}
\epsfig{angle=270,file=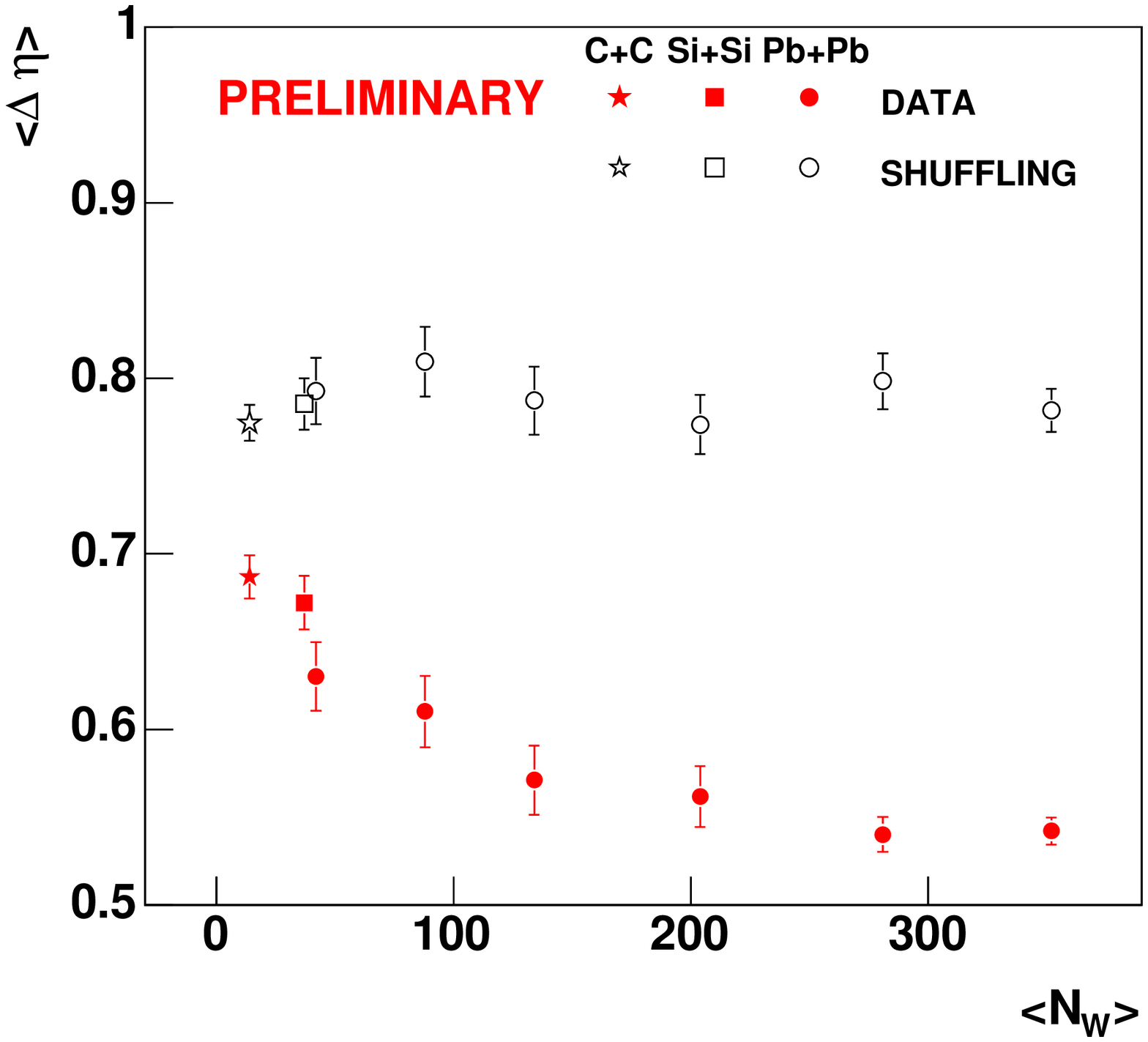,width=6.5cm}
\end{center}
\caption{The dependence of the BF 's width on the number of
wounded nucleons for p+p, C+C, Si+Si and Pb+Pb collisions at
$\sqrt{s_{NN}} = 17.2$ GeV (left plot) and at $\sqrt{s_{NN}} = 8.8$ 
GeV (right plot). } \label{centrality1}
\end{figure}

Fig. \ref{centrality1} shows the dependence of the width $\langle \Delta \eta \rangle$
of the BF on the mean number of wounded nucleons \cite{BF_NA49}. We notice a strong 
centrality dependence of the width $\langle \Delta \eta \rangle$, with a maximum value 
for p+p collisions. Furthermore, the width $\langle \Delta \eta \rangle$ decreases as 
we move from the most peripheral to the most central Pb+Pb collisions by an amount of 
$17 \%$. On the other hand the width of the BF from both HIJING and shuffled data 
does not show any clear dependence on centrality.

The results from a similar analysis by the STAR collaboration at 
RHIC \cite{STAR}, demonstrate that the width of the BF for Au + Au 
collisions at $\sqrt{s_{NN}} = 130$ GeV also decreases monotonically in a similar 
manner. This decrease was taken as an indication of a change in the particle production 
dynamics \cite{STAR}. The decrease in the NA49 data is of the order of $17 \pm 3\%$, 
whereas for the higher energy STAR data it is of the order of $14 \pm 2\%$. 

The measured narrowing of the BF is qualitatively consistent with the delayed 
hadronization scenario \cite{Pratt,STAR} of an initially deconfined phase. 
Several model calculations have been published which provide a more quantitative 
description \cite{Statistical,Resonance1,Resonance2,Bialas}. Hadronization was 
modelled in \cite{Statistical,Resonance1,Resonance2} by a thermal fireball model 
and by incorporating the experimentally observed strong radial flow. Using only 
a single fireball \cite{Resonance1,Resonance2} it was not possible to reproduce 
the measured degree of narrowing of the BF with increasing centrality. On the 
other hand, assumption of several smaller fireballs with individual charge 
conservation provides a reasonable description of the STAR results \cite{Statistical}. 
The quark coalescence model was applied to the hadronization of the deconfined 
phase in \cite{Bialas}. 

Finally, Fig. \ref{centrality1} shows also the preliminary results on the dependence 
of the width $\langle \Delta \eta \rangle$ of the BF on the mean number of wounded 
nucleons for $\sqrt{s_{NN}} = 8.8$ GeV. Note that these results are obtained without 
use of the acceptance filter. We notice a strong centrality dependence of the width, 
with a maximum value for C+C collisions. Furthermore, the width decreases as we move 
from the most peripheral to the most central Pb+Pb collisions by an amount of $14 \pm 3\%$. 
On the other hand the width of the BF from shuffled data does not show any clear 
dependence on centrality. 

\vspace{0.2 cm}

%%%%%%%%%%%%%%%%%%%%%%%%%%%%%%%%%%%%%
\noindent \textbf{3. Summary}
%%%%%%%%%%%%%%%%%%%%%%%%%%%%%%%%%%%%%

In this contribution, we have presented measurements of the BF, in p+p, C+C and Si+Si 
collisions as well as in centrality selected Pb+Pb collisions at $\sqrt{s_{NN}} = 17.2$ 
and $\sqrt{s_{NN}} = 8.8$ GeV. The results show a strong system size and centrality 
dependence of the width. The width decreases monotonically from peripheral to  central 
Pb+Pb collisions. This decrease is of the order of $17 \pm 3\%$ and $14 \pm 3\%$ 
respectivelly. On the other hand, the width of the BF coming from shuffled data does 
not show any sign of dependence on centrality.
 
Our results, even if they correspond to different phase space and acceptance filter, 
agree qualitatively with those of the STAR experiment which have shown the same centrality dependence of the width of the BF at higher collision energy, with a decrease of the 
order of $14 \pm 2\%$.

\vspace{0.2 cm}

\noindent \textbf{Acknowledgements:} 
We would like to thank Dr. F. Diakonos for his helpful discussions during the 
first stage of this work.

\end{document}